\documentstyle[12pt]{article}
\makeatletter
\ifcase\@ptsize
 \font\teneufm=eufm10
 \font\seveneufm=eufm7
 \font\fiveeufm=eufm5
 \font\teneusm=eusm10
 \font\seveneusm=eusm7
 \font\fiveeusm=eusm5
\or
 \font\teneufm=eufm10 scaled \magstephalf
 \font\seveneufm=eufm7
 \font\fiveeufm=eufm5
 \font\teneusm=eusm10 scaled \magstephalf
 \font\seveneusm=eusm7
 \font\fiveeusm=eusm5
\or
 \font\teneufm=eufm10 scaled \magstep1
 \font\seveneufm=eufm7
 \font\fiveeufm=eufm5
 \font\teneusm=eusm10 scaled \magstep1
 \font\seveneusm=eusm7
 \font\fiveeusm=eusm5
\fi

\newfam\eufmfam
\newfam\eusmfam
\textfont\eufmfam=\teneufm  \scriptfont\eufmfam=\seveneufm
  \scriptscriptfont\eufmfam=\fiveeufm
\textfont\eusmfam=\teneusm  \scriptfont\eusmfam=\seveneusm
  \scriptscriptfont\eusmfam=\fiveeusm

\def\frak{\ifmmode\let\next\frak@\else
 \def\next{\errmessage{Use \string\frak\space only in math mode}}\fi\next}
\def\frak@#1{{\frak@@{#1}}}
\def\frak@@#1{\fam\eufmfam#1}
\let\goth\frak
\def\sh{\ifmmode\let\next\sh@\else
 \def\next{\errmessage{Use \string\sh\space only in math mode}}\fi\next}
\def\sh@#1{{\sh@@{#1}}}
\def\sh@@#1{\fam\eusmfam#1}

\ifcase\@ptsize
 \font\tenmsa=msam10
 \font\sevenmsa=msam7
 \font\fivemsa=msam5
 \font\tenmsb=msbm10
 \font\sevenmsb=msbm7
 \font\fivemsb=msbm5
\or
 \font\tenmsa=msam10 scaled \magstephalf
 \font\sevenmsa=msam7
 \font\fivemsa=msam5
 \font\tenmsb=msbm10 scaled \magstephalf
 \font\sevenmsb=msbm7
 \font\fivemsb=msbm5
\or
 \font\tenmsa=msam10 scaled \magstep1
 \font\sevenmsa=msam7
 \font\fivemsa=msam5
 \font\tenmsb=msbm10 scaled \magstep1
 \font\sevenmsb=msbm7
 \font\fivemsb=msbm5
\fi

\newfam\msafam
\newfam\msbfam
\textfont\msafam=\tenmsa  \scriptfont\msafam=\sevenmsa
  \scriptscriptfont\msafam=\fivemsa
\textfont\msbfam=\tenmsb  \scriptfont\msbfam=\sevenmsb
  \scriptscriptfont\msbfam=\fivemsb

\def\Bbb{\ifmmode\let\next\Bbb@\else
 \def\next{\errmessage{Use \string\Bbb\space only in math mode}}\fi\next}
\def\Bbb@#1{{\Bbb@@{#1}}}
\def\Bbb@@#1{\fam\msbfam#1}
\def\hexnumber@#1{\ifnum#1<10 \number#1\else
 \ifnum#1=10 A\else\ifnum#1=11 B\else\ifnum#1=12 C\else
 \ifnum#1=13 D\else\ifnum#1=14 E\else\ifnum#1=15 F\fi\fi\fi\fi\fi\fi\fi}
\def\msa@{\hexnumber@\msafam}
\def\msb@{\hexnumber@\msbfam}
\mathchardef\square="0\msa@03

\makeatother
\newcommand{\beq}{\begin{equation}}
\newcommand{\eeq}{\end{equation}}
\newcommand{\ba}{\begin{array}}
\newcommand{\ea}{\end{array}}
\newcommand{\bea}{\begin{eqnarray}}
\newcommand{\eea}{\end{eqnarray}}
\newcommand{\bean}{\begin{eqnarray*}}
\newcommand{\eean}{\end{eqnarray*}}
\newtheorem{theorem}{Theorem}[section]
\newtheorem{prop}[theorem]{Proposition}
\newtheorem{lem}[theorem]{Lemma}
\newtheorem{defi}[theorem]{Definition}
\newtheorem{remark}[theorem]{Remark}
\newenvironment{rem}{\begin{remark} \rm}{\end{remark}}
\newtheorem{proof}{Proof.}
\newenvironment{pf}{\begin{proof} \rm}{\hfill$\square$ \end{proof}}

\makeatletter
\@addtoreset{equation}{section}

\makeatother

\newcommand{\RR}{{\Bbb R}}
\newcommand{\CC}{{\Bbb C}}
\newcommand{\PP}{{\Bbb P}}

\newcommand{\DD}{{\Bbb D}}
\def\th{\theta}         
         \def\Ga{\Gamma}
\def\be{\beta}
\def\al{\alpha}

\def\sig{\sigma}        \def\Sig{\Sigma}

\newcommand{\ajm}[3]{Amer. J. Math. {\bf #1} (#2), #3}
\newcommand{\cmp}[3]{Comm. Math. Phys. {\bf #1} (#2), #3}
\newcommand{\pl}[3]{Phys. Lett. {\bf B #1} (#2) #3}
\newcommand{\np}[3]{Nucl. Phys. {\bf B #1} (#2), #3}

\newcommand{\ihes}[3]{Publ. Math. I.H.E.S. {\bf #1} (#2), #3}

\newcommand{\ijmp}[3]{Int. Jour. Mod. Phys. A{\bf #1} (#2), #3}

\newcommand{\lmp}[3]{Lett. Math. Phys. {\bf #1} (#2), #3}
\newcommand{\plms}[3]{Proc. London Math. Soc. {\bf #1} (#2), #3}

\newcommand{\tams}[3]{Trans. AMS {\bf #1} (#2), #3}
\newcommand{\lanl}[1]{LANL preprint, hep-th/#1}
\newcommand{\rref}[1]{(\ref{#1})} 
\newcommand{\half}{\frac{1}{2}}
\newcommand{\dz}{\mbox{d}z}
\newcommand{\dzb}{\mbox{d}\bar z}
\newcommand{\newblock}{{}}

\def\dsl{\displaystyle}

\def\RS {Riemann surface}\def\RSs {Riemann surfaces}
\newcommand{\VHS}{Variation of Hodge Structure}
\newcommand{\VHSs}{Variations of Hodge Structure}
\def\deriv#1#2{{\strut{\del#1\over\del#2}\displaystyle}}
\def\plusminus{\pm}

\newcommand{\del}{{\partial}}
\newcommand{\delb}{{\bar\partial}}

\begin{document}
\begin{titlepage}
\begin{flushright}{\AA}rhus Universitet\\
Matematisk Institut\\ Preprint series 1993 No. 32
\\[7pt]
Preprint Montpellier PM/93 -- 43\\[7pt]
hep-th/9312093
\end{flushright}
\vspace{0.8truecm}
\begin{center}
{\huge Geometry of Higgs and Toda \\ \vspace{0.2truecm}
Fields on Riemann Surfaces}
\end{center}
\vspace{0.8truecm}
\begin{center}
{\large\it Ettore Aldrovandi} \footnote{Supported
by the Danish Natural Science Research
Council.}${}^,$\footnote{e--mail: ettore@mi.aau.dk}\\
Matematisk Institut -- \AA rhus Universitet\\
Ny Munkegade -- DK-8000 \AA rhus C --- DANMARK
\\[0.5truecm]
{\large\it Gregorio Falqui}
\footnote{e--mail: falqui@lpm.univ-montp2.fr}\\
Laboratoire de Physique Math\'ematique \footnote{Unit\'e de recherche
associ\'ee au CNRS n. 768} ---
Universit\'e Montpellier II \\
Place E. Bataillon, 34095 Montpellier CEDEX 05 --- FRANCE
\end{center}
\vspace{0.2truecm}
\begin{center}
December 1993. Revised, June 1994.
\end{center}
\vspace{0.2truecm}
\abstract{\noindent
We discuss geometrical aspects of Higgs systems and Toda field
theory in the framework of the theory of vector bundles on
Riemann surfaces of genus greater than one. We point out how Toda fields
can be considered as equivalent to Higgs systems -- a connection on
a vector bundle $E$ together with an End($E$)--valued one form both
in the standard and in the Conformal Affine case.
We discuss how variations of Hodge
structures can arise
 in such a framework and determine holomorphic embeddings of Riemann surfaces
into locally homogeneous spaces, thus giving hints to
possible
realizations of $W_n$--geometries. }
\end{titlepage}
\setcounter{footnote}{0}

\section{Introduction}
The amount of research activity devoted to the study of conformal
and integrable systems in two dimensions has reached a considerably
high rate in the last years. In particular, great attention
has been paid to Toda field theories and extended conformal (or $W$)
symmetries, and the efforts done in this direction starting from
the pioneering papers of Zamolodchikov~\cite{Za85},
Gervais and Neveu~\cite{GeNeYY} and Fateev
and Lukyanov~\cite{FaLu88} have attained beautiful results both in the
classical and in the quantum case.

On a more general level, one of the most remarkable achievements
can be also considered
bringing to the light
the richness of the mathematical structure underlying
such theories, and the deep relationships existing between
{\em a priori} different field theoretical models.

In this paper we want to give a contribution in such a direction,
and namely in the study of some outcomes of the links between the Toda
equations and the geometry of Higgs bundles or Higgs
pairs, in the framework of the theory of (analytic) vector bundles
(and connections thereupon) on \RSs\  of generic genus.

A Higgs bundle is a system
composed of a connection $A$ on a vector bundle $E$ over a
\RS\  $\Sigma$ and
a holomorphic endomorphism  $\theta$ of $E$ satisfying
\beq \label{histr}F_A+[\theta
,\theta^*]=0\eeq
Such structures
were first introduced by Hitchin~\cite{Hit87},
in the framework of self--dual Yang Mills equations.
The same author, in
\cite{Hit87b}, proved their relevance
showing that they consitute a remarkable example of an algebraically
completely
integrable hamiltonian system. Later on, a growing number of
publications
were devoted to the study of applications of Higgs
pairs in the theory of harmonic bundles,
local systems, uniformization and Variations
of Hodge Structures (see, e.g.,~\cite{Cor88,Simp88,Simp92}).

Our starting point is the zero--curvature representations of the Toda
equations, (both in the standard and in the Conformal Affine cases)
which in a suitable gauge (see also~\cite{GeO'RRaSa})
can be seen to be equivalent to Hitchin's equation (\ref{histr}) for
the corresponding Higgs pair. Although it is not a difficult outcome
of the Toda equations,
{\em this is one of the central points of the paper}.

We can then proceed
in two directions. At first we can adapt
some nowadays fairly standard
computations in Toda Field theory~\cite{BiGe89} to prove that
the $A_{n-1}$--Toda connection, which
is naturally defined on the direct sum
$E = \bigoplus_{r = 0}^{n-1} K^{-\frac{n-1}{2}+r}$ of powers of the
canonical line bundle $K$ on $\Sigma ,$  is mapped to an
{\em analytic} flat connection on the full $(n-1)^{th}$--jet
extension $V=J(K^{-\frac{n-1}{2}})$ of $ K^{-\frac{n-1}{2}}$, whose degrees
of freedom are parameterized by $W_n$--fields.

On the other side, it is possible to decompose the Toda connection
in a {\em metric} part plus a deformation $\alpha$, which is simply the
sum of
the Higgs field and its metric conjugate. The structure equations
for the Higgs pair translate into a harmonicity condition for
the one--form $\al$. This means that,
associated to a Toda system, there is a natural harmonic twisted map
$f_H$
from $\Sigma$ to an homogeneous manifold. Furthermore,
the Toda connection (in the standard case),
can be shown to satisfy the so--called
Griffiths transversality conditions~\cite{Gri70,Simp88} and so defines
a variation of a Hodge structure, a fact already noticed in~\cite{CeVa91} in
the framework of $N=2$ superconformal models and their integrable
deformations.
This entails that the map $f_H$ is
actually a {\em holomorphic embedding} of $\Sigma$ into the quotient
$\Gamma \backslash{\Bbb D}$ of a Griffiths period domain ${\Bbb D}$ by the
 monodromy group $\Gamma$.

Henceforth, through the theory of Higgs bundles we can associate to a
Toda field (and, by the discussion above, to a $W_n$--algebra)
a holomorphic
map of $\Sigma$ into a hermitean manifold,
a detour which can be also suggested by the fact that the
construction of the
Poisson commuting conserved quantities in~\cite{Hit87b}
is reflected in the definition
of $W_n$ algebras via higher order Casimir invariants~\cite{BaBoScSu88},
and is under current investigation in the framework of Langlands--Drinfel'd
correspondences (see, e.g.~\cite{Fr91}, and the references quoted therein).

It is natural thus to interpret such structures as a possible realizations
of the $W_n$ geometries, introduced in~\cite{SoSta91,GeMa92} (and
more recently discussed in the framework of BRST symmetries in~\cite{Zu92}).
Although we do not perform here a thorough
comparison between our framework and such results,
we will make some comments about the relationship
between our picture and the latest results of
Gervais, Razumov and Saveliev~\cite{RaSa93,GeSa93} about $W_n$ geometry
and
generalized Pl\"ucker embeddings associated to Toda systems (Section 6).

Let us  sketch the plan of the paper. In Section 2 we collect
some informations about Higgs pairs and harmonic
bundles from~\cite{Hit87,Cor88,Simp92}. Then, in section 3
we recall how the Toda equations can be seen
as a zero--curvature condition, and describe the abovementioned
equivalence between Toda fields and Higgs pairs. The extension to the
Conformal Affine case is also discussed. We devote Section 4 to the
description of the mapping between Toda bundles and $W_n$
bundles on curves of arbitrary genus, pointing out their global
aspects.  In Section 5, after reviewing the basics of Griffiths
theory of Variations of Hodge Structures, we show how the Toda equations
fit into this framework and prove the holomorphicity
of the embedding of $\Sigma$ determined by the metric part of the
Toda connection. We finally put our observations and comments in
Section 6.

\section{Higgs systems and harmonic bundles on a \protect\RS}
\label{higgs}

Let $\Sigma$ be a genus $g$ \RS\
with canonical bundle $K$.
\begin{defi}
A Higgs bundle over $\Sigma$ is a pair $(E,\theta)$ with E a holomorphic
vector bundle over $\Sigma$ and $\theta\in H^0(\mbox{End}\, E\otimes K)$
\par A Higgs bundle is stable if Mumford's inequality
\begin{equation}
\dsl{{c_1(F)\over {\mbox{rk} F}}<{c_1(E)\over {\mbox{rk} E}}}
\end{equation}
holds for every non--trivial {\em $\theta$ -- invariant}
subbundle $F\subset E$.
\end{defi}
The generalization of Narasimhan -- Seshadri theorem holds in the
following form~\cite{Hit87,Simp88}
\begin{theorem}
If $(E,\theta)$ is stable and $c_1(E) = 0$ there is a unique unitary
connection $\nabla_H$ compatible with the holomorphic structure, such
that
\begin{equation}\label{nigel}
F_H +[\theta,\theta^*]=0
\end{equation}
\end{theorem}
The basic properties of Higgs systems we are going to use in the sequel
are the following.

Given a connection $\nabla_H$ whose curvature
equals the commutator
$[\theta,\theta^*]$ for a holomorphic section
$\theta\in H^0(\mbox{End} E\otimes K)$ then
\begin{equation}
\nabla = \nabla_H + \theta +\theta^*
\label{flat}
\end{equation}
is a {\em flat} $GL(n,{\Bbb C})$--connection, where
$\rho (X)=- X^*$ is the anti--involution defining the compact real
form of the group.

In~\cite{Hit92} the holonomy of $\nabla$ and $\nabla_H$ are
characterized following the arguments below. Let ${\frak g}^{\Bbb C}$
be a simple Lie algebra and let ${\frak h}=\{h_1,e_1,f_1\}$ be a
principal $sl_2$ subalgebra. Let $e_1,\ldots,e_k$ highest weight vectors
of the irreducible representations in which ${\frak g}^{\Bbb C}$ is
branched under ${\frak h}$. Then there exists a Lie algebra involution
$\sigma$ of ${\frak g}^{\Bbb C}$ sending $f_1\to -f_1$ and $e_i\to -e_i
,\quad i=1,\ldots,k$. The fixed point set of $\sigma$ turns out to be
the complexification of a maximal compact subalgebra of the split
real form ${\frak g}^{r}$ of ${\frak g}^{\Bbb C}$. \par Since $\sigma$
commutes with $\rho$ a careful application of Lie algebra
properties along the lines of~\cite{Kos59}
proves that $\nabla_H$ has holonomy contained in the
maximal compact subalgebra of ${\frak g}^{\Bbb C}$, while the flat
connection $\nabla = \nabla_H + \theta +\theta^*$ has holonomy contained
in ${\frak g}^{r}$.

The notion of Higgs system can be related to the one of {\em harmonic
bundle} in a way we will briefly illustrate. A Higgs system $(E,\th )$
such that \rref{nigel} is satisfied clearly defines a pair
$(V,\nabla )$, where $V$ is the $C^\infty$ bundle underlying
$E$ and $\nabla$ is
the flat connection \rref{flat}. Let us now consider a
complex rank $n$ vector bundle
$V$ equipped with a flat connection $\nabla$. As it is well known,
the introduction of an hermitean
fibre metric $H$ on $V$ amounts to a reduction of the structure group from
$GL(n,\CC)$ to $U(n)$, and allows for a splitting
\begin{equation}
\nabla = \nabla_{\! H} + \al
\label{flat2}
\end{equation}
where $\nabla_H$ is a unitary connection and $\al$ is a ($C^\infty$)
1--form with values in (the self--adjoint part of) $\mbox{End}(V)$.
Clearly, the connection $\nabla$ being flat is equivalent to the
following pair of equations \cite{Cor88}
\begin{eqnarray}
& &\nabla_H^2 + \half [\al , \al ] = 0\label{cor1}\\
& &\nabla_{\! H}\al = 0\label{cor2}
\end{eqnarray}

\begin{defi}[Corlette, Donaldson, Simpson]
We define the pair $(V\, ,\nabla )$ to be {\em harmonic}
or equivalently speak of a {\em harmonic metric} $H$ if, under the
splitting \rref{flat2}, we have
\begin{equation}
{\nabla_{\! H}}^{\! *}\,\al = 0
\label{harmonic}
\end{equation}
where the adjoint is taken with respect to a given metric on $\Sig$.
\end{defi}

It is then a comparatively easy, but nontheless important remark that
since $\al$ is self--adjoint, we can decompose it
as $\al = \th + \th^*$, thus showing that equations
(\ref{cor1}, \ref{cor2}, \ref{harmonic}) are equivalent to Hitchin's
self--duality equation \rref{nigel}, supplemented by $\delb\,\th =
0$\footnote{The $\delb$--operator clearly comes from the (0,1) part of
$\nabla_H$} \cite{Don87,Simp92}.

The reason why a bundle or a metric satisfying (\ref{cor1},
\ref{cor2}, \ref{harmonic}) are called harmonic is to
be understood in the following sense~\cite{Cor88}. A metric $H$
can be considered as a multivalued mapping $f_H:\Sig\rightarrow
GL(n,\CC)/U(n)$ or, in other words, as a section of a bundle over
$\Sig$ whose standard fibre is the coset $GL(n,\CC)/U(n)$. Since
$\nabla$ is flat, $V$ itself and all its associated bundles come from
a representation of the fundamental group of $\Sig$. The
section $f_H$ can in fact be regarded as a map from
the universal cover of $\Sig$,
$f_H:\widetilde{\Sig}\rightarrow GL(n,\CC)/U(n)$, equivariant with
respect to the action of $\pi_1(\Sig )$. In other words we have the
commutative diagram
\[
\setlength{\unitlength}{0.0070in}%
\begin{picture}(260,113)(160,540)
\put(175,545){\vector( 1, 0){163}}
\put(430,630){\vector( 0,-1){ 65}}
\put(160,630){\vector( 0,-1){ 65}}
\put(175,645){\vector( 1, 0){163}}
\put(150,540){\makebox(0,0)[lb]{\raisebox{0pt}[0pt][0pt]{$\Sigma$}}}
\put(350,540){\makebox(0,0)[lb]{\raisebox{0pt}[0pt][0pt]
{$\Gamma\backslash GL(n,\CC)/U(n)$}}}
\put(250,555){\makebox(0,0)[lb]{\raisebox{0pt}[0pt][0pt]
{$\scriptstyle{f_H}$}}}
\put(150,640){\makebox(0,0)[lb]
{\raisebox{0pt}[0pt][0pt]{$\widetilde{\Sigma}$}}}
\put(350,640){\makebox(0,0)[lb]{\raisebox{0pt}[0pt][0pt]
{$GL(n,\CC)/U(n)$}}}
\put(250,655){\makebox(0,0)[lb]{\raisebox{0pt}[0pt][0pt]
{$\scriptstyle{f_H}$}}}
\end{picture}
\]
Here $\pi_1(\Sig )$ acts on $\widetilde{\Sigma}$ as the group of
deck transformations and on $GL(n,\CC)$ via the holonomy representation
$\Gamma$.
It is well known that there  exists a flat
coordinate system for $V$ in which $\nabla$ is simply given by the
exterior differential $d$. In these coordinates one has
\begin{equation}
\al = -\half f_H^{-1}\, d f_H
\label{alpha}
\end{equation}
which means that $\al$ can be identified with the differential of
$f_H$, and therefore equation \rref{harmonic} implies that the map $f_H$
is harmonic, according to the Eells--Sampson characterization of
harmonic maps \cite{EeSam64}. See \cite{Don87,Simp92} for details.
We shall refer to the map $f_H$ so obtained as the ``classifying map''
and by a slight abuse of language, a ``harmonic bundle'' will mean either
the Higgs system $(E,\th)$ satisfying \rref{nigel} or the related
$C^\infty$ pair $(V,\nabla )$.

Thus what we are going to consider in the sequel are harmonic bundles
{\em plus\/} the additional structure given by the reduction of the
structure group to a split real form~\cite{Hit92}.

Let us consider the bundle
\begin{equation}
E = \bigoplus_{r = 0}^{n-1} K^{-\frac{n-1}{2}+r}
\label{E:direct}
\end{equation}
Its determinant is trivial, therefore we can consider
its structure group to be the semisimple group $G^\CC
=SL(n,\CC)$. According to \cite{Hit92}, we take $\th$ to be given by
\begin{equation}
\th =
\left(\begin{array}{ccccc}
     0&     1&      &\cdots&0 \\
   u_2&     0&     1&      &0 \\
   u_3&\ddots&\ddots&\ddots&0 \\
\vdots&\ddots&\ddots&     0&1 \\
   u_n&\cdots&   u_3&   u_2&0\end{array}\right)
\label{thetabig}
\end{equation}
where $u_r\in H^0(\Sig ,K^r)$ for $r=2,\dots ,n$.
It follows from \cite{Simp92}, Lemmas 2.11 and 2.12, that the holonomy
representation given by the pair $(E,\th)$ is defined over $\RR$ if
and only if there exist a symmetric bilinear map $S:E\otimes
E\rightarrow{\cal O}_\Sig$ satisfying
\[
S(\th u,v)=S(u,\th v)
\]
for any two local sections $u,v$ of $E$. This happens to be the case
with the pair defined by \rref{E:direct} and \rref{thetabig}, and with
the map $S$ given by the matrix \cite{DeBGoe}
\begin{equation}
S=\left(\begin{array}{ccccc}
  & & & &1 \\
  & & &\cdot& \\
  & &\cdot& & \\
  &\cdot& & & \\
 1& & & &  \end{array}\right)\label{S}
\end{equation}
Of course, this only tells that the structure group,
and hence the holonomy, is reduced to a real form of
$G^\CC = SL(n,\CC)$. Then Hitchin's analysis briefly resumed earlier tells
that this
is in fact the {\em split\/} form $G^r =SL(n,\RR)$. It is of some
interest to notice that the real form $SL(n,\RR)$ appears here in a
rather disguised form, that is, the conjugation $\tau$ in $sl(n,\CC)$
which selects the split real form is concretely given by
\begin{equation}
\tau (\xi )= S\,\bar\xi S,\qquad\xi\in sl(n,\CC)\label{tau}
\end{equation}
which nevertheless can be shown to be conjugate to the standard split
real form given by $\xi\rightarrow\bar\xi$.

\section{The Toda equations and their link with Higgs bundles}

Let ${\frak g}$ be a simple finite dimensional Lie algebra and let
us consider a Cartan -- Weyl basis
\begin{eqnarray*}
\left[h_i,h_j\right] &=&  0 \\
\left[h_i,e_{\pm\alpha}\right] &=& \pm \alpha_i e_{\pm\alpha} \\
\left[e_{\alpha},e_{-\alpha}\right] &=& \sum \alpha_i h_i
\end{eqnarray*}

A Toda field over a Riemann surface $\Sigma$ is a field $\Phi$
taking values
in the Cartan subalgebra of ${\frak g}$ and satisfying the equations
\begin{equation}\label{toda}
\del_z\del_{\bar z}\Phi=\sum h_i \mbox{e}^{\alpha_i(\Phi) }
\end{equation}
It is a well known fact that the equations (\ref{toda}) can be obtained
as the compatibility condition for a linear system~\cite{LeSa}. Let us
rederive this result in the framework of the theory of connections.\\
Let us denote with $\Delta_+^s$ ($\Delta_-^s$)
the set of all positive (negative)
simple roots and with ${\cal E}_+$ (${\cal E}_-$) their sum,
and define a ${\frak g}$--valued local 1--form $A=A_z \mbox{d}z+ A_{\bar z}
\mbox{d}{\bar z}$ as
\begin{eqnarray}
A_z &=& \half \del_z\Phi + \exp (\half \mbox{ad}\Phi)\cdot {\cal E}_+\\
A_{\bar z} &=& -\half \del_{\bar z}\Phi +\exp (-\half
\mbox{ad}\Phi)\cdot {\cal E}_-
\end{eqnarray}
The curvature $F_A$ is a $(1,1)$--form
\begin{displaymath}
F_{z\,\bar z}=\del_{ z}A_{\bar z} - \del_{\bar z} A_z + [A_z,A_{\bar z}]
\end{displaymath}
Hence we get
\begin{displaymath}
F_{z\,\bar z}=-\del_z\del_{\bar z}\Phi+
[\exp(\half \mbox{ad}\Phi)\cdot {\cal E}_+\, ,\,
\exp (-\half \mbox{ad}\Phi)\cdot {\cal E}_-]
\end{displaymath}
Since $[\Phi,E_\pm]=\pm\sum_{\alpha\in\Delta^s_+}\alpha(\Phi)
e_{\plusminus\alpha}$,
we have
\[
\exp(\pm \half \mbox{ad}\Phi)\cdot {\cal E}_\pm=\sum_{\alpha\in\Delta^s_+}
\exp \half\alpha(\Phi)\, e_{\pm\alpha}
\]
and so
\bean
F_{z\, \bar z} &=& -\del_z\del_{\bar z}\Phi+
\sum_{\alpha,\beta\in\Delta^s_+}\exp (\half\alpha(\Phi)+\beta(\Phi))
\left[e_{\alpha},e_\beta\right]\\
&=& -\del_z\del_{\bar z}\Phi + \sum h_i\exp(\alpha_i(\Phi))
\eean
Let us now consider the gauge transformed connection under the element
$g=\exp(\half\Phi)$ \cite{bab89}. We have
\begin{eqnarray}
& &A_z^g = \del_z\Phi + {\cal E}_+\label{T:gauge-1}\\
& &A_{\bar z}^g = \exp (- \mbox{ad}\Phi)\cdot {\cal E}_-\label{T:gauge-2}
\end{eqnarray}
This form leads directly to the Higgs bundle picture. In fact, we can
consider $\exp(\Phi)$ as an hermitean form on the fibers and split
$D_A=d+A$ as
\begin{equation}
D_H+\theta+\tilde\theta
\end{equation}
where $D_H$ is the metric connection associated to $H=\exp(\Phi)$,
\begin{equation}
\theta={\cal E}_+ \mbox{d}z
\label{theta}
\end{equation}
and $\tilde\theta=H^{-1} {\cal E}_- H$.
Namely we have that
\begin{eqnarray*}
{D_H}^\prime &=& \del+\del\Phi\\
{D_H}^{\prime\prime} &=& \bar\del
\end{eqnarray*}
Since the conjugation $\rho$ acts as
$\rho (h_i)=-h_i\, ,\; \rho (e^+_i)=-e^-_i$,
we get that
$\tilde\theta$ is the metric adjoint endomorphism of $\theta$. Hence,
together with the obvious fact that  ${D_H}^{\prime\prime}\theta=0$,
the zero--curvature equations in this gauge are
\begin{equation}
{D_H}^2+\left[\theta,\theta^*\right]=0
\end{equation}
thus showing that
any solution to the Toda equations gives rise to a well defined
solution of the Hitchin equations for the Higgs bundle. We follow
\cite{GeO'RRaSa} and call {\em Toda--gauge\/} the one  where the
connection takes the form \rref{T:gauge-1}, \rref{T:gauge-2}.

The endomorphism $\th$ constructed above
correspond exactly to the one given by \rref{thetabig} if we set all
the $u_r$'s to zero, while the vector bundle $E$ is given by
\rref{E:direct}, where $n-1$ is the rank of the Lie algebra, i.e.
${\frak g}^\CC=A_{n-1}$.  The metric $H$ is given by a diagonal matrix
whose entries $h_r=e^{\varphi_r}\, ,\; r=1,\dots ,n$, are themselves
metrics on the factors $K^{-\frac{n-1}{2}+r-1}$ appearing in
\rref{E:direct}. This completely fixes the transformation law of the
fields $\varphi_r$, and one can check it coincides with the
well--known conformal transformation properties of the Toda fields
\cite{bab89}.

We wish now to extend the correspondence between Toda--like models
and Higgs systems to the so--called Conformal Affine Toda models
\cite{BaBo90}.
To this purpose, we retain the same form for the metric
$H$ but modify the endomorphism $\th$ to
\begin{equation}
\th = {\cal E}_+ + u\, e_{-\psi}
\label{theta:aff}
\end{equation}
where $\psi$ is the longest root and $e_{\pm\psi}$ are the positive
and negative root vectors.
It follows that $\th^*$ will be given by
\begin{equation}
\th^*= \exp\, (-\mbox{ad}\Phi )\cdot ({\cal E}_- + \bar u\, e_{\psi})
\label{theta*:aff}
\end{equation}
(notice that ${\cal E}_- + \bar u\, e_{\psi} = -\rho ({\cal E}_+ + u\,
e_{-\psi})$). Since $\psi$ is the longest root, the element
$e_{-\psi}$ has degree $-k\equiv -\mbox{rank}{\frak g}$ with respect
to the action of the
principal $sl_2$ subalgebra $\frak h$~\cite{Kos59} and therefore we
must have $u\in H^0(\Sig\, ,\,  K^{k+1})$ \cite{Hit92}, which in the
$A_{n-1}$--case, for instance, means that $u$ is a $(n,0)$--weight
differential, i.e. a section of ${(T_\Sig^*)}^{\otimes\, n}\equiv K^n$.

To clarify the procedure, let us decompose
\[ A = B + \eta \]
where $\eta$ is the 1--form
\bean
\eta &=& u\, e_{-\psi}\,\mbox{d}z + \bar u\,
e^{-\mbox{\tiny ad}\Phi}e_\psi\,\mbox{d}\bar z \\
&=& u\, e_{-\psi}\,\mbox{d}z + \bar u\, e^{-\psi (\Phi
)}e_\psi\,\mbox{d}\bar z
\eean
in such a way that $B$ is the connection associated to the standard Toda
equations. Using the following fairly standard relation
\[
F_A = F_B + \half [\eta ,\eta ] + D_B\eta
\]
and $[{\cal E}_+,e_\psi ]=[e^{-\mbox{\tiny ad}\Phi}{\cal E}_-,e_{-\psi}]=0$,
because $\psi$ is the longest root, we calculate
\bean
\half [\eta ,\eta ] &=& -|u|^2\, e^{-\psi (\Phi )}\, [e_\psi
,e_{-\psi}]\,\dz\wedge\dzb\\
D_B\eta &=& \left( \overline{\del_{\bar z}u}\, e^{-\psi (\Phi )}e_\psi
- \del_{\bar z}u\, e_{-\psi}\right)\,\dz\wedge\dzb
\eean
so that looking at the generators, we see that the equation $F_A=0$
implies
\bean
& &F_B + \half [\eta ,\eta ] = 0\\
& &D_B\eta = 0
\eean
separately, which gives
\bea
\del_z\del_{\bar z}\Phi &=& \sum h_i e^{\alpha_i(\Phi)} - |u|^2\,
e^{-\psi (\Phi )}\,h_\psi\label{CAT1:1} \\
\delb u &=& 0\label{CAT1:2}
\eea
where $h_\psi = [e_\psi ,e_{-\psi}]$. Now, let us locally put $|u|^2 =
e^{2\eta}$ so that $\delb u = 0$ yields $\del\delb\eta = 0$. Then
equations \rref{CAT1:1} and \rref{CAT1:2} read
\bea
\del_z\del_{\bar z}\Phi &=& \sum h_i e^{\alpha_i(\Phi)} -
e^{2\eta -\psi (\Phi )}\,h_\psi\label{CAT2:1}\\
\del_z\del_{\bar z}\eta &=& 0\label{CAT2:2}
\eea
which essentially coincide with the equations defining the ``Conformal
Affine Toda'' model~\cite{BaBo90,AFGZ91}. Actually,
in our formulation the field associated to the
extra central generator in the affine algebra is missing.
However this is not a serious problem, as the dynamics of this missing
field is completely fixed by the other ones, whose equations of motion
are correctly reproduced. We wish to stress that the conformal
invariance of equations
\rref{CAT2:1}, \rref{CAT2:2} naturally arises in the present setting,
as they are interpreted as the integrability condition for a
connection on a globally well-defined vector bundle.

We can summarize
the results of this section in the following
\begin{prop}
A solution of the standard Toda equations \rref{toda}
gives rise to a well defined solution
of the Hitchin's equation \rref{nigel} whose underlying Higgs system
is given by \rref{theta} above. The same
statement applies to the Conformal Affine Toda equations
\rref{CAT2:1}, \rref{CAT2:2}, with \rref{theta} replaced by
\rref{theta:aff}.
\end{prop}

Let us remark that the above
set--up allows to interpret the
limit from Conformal Affine Toda to standard Toda (see \cite{BaBo90})
in a nice geometrical fashion.
In fact, according to \cite{Hit92}, we can regard the former as a
``deformation'' of the standard  Toda model related to a deformation
of the associated Higgs bundle.

\section{Toda systems and $W_n$ algebras}
\label{W-section}

In this section we will make explicit some aspects of the relations
between Toda equations and $W_n$ algebras~\cite{GeNeYY,BiGe89} through the
theory of connections. It has already been established
(see, e.g.~\cite{FaLu88,BiFoKo92})
that classical $W_n$ algebras can be associated to the
Drinfel'd--Sokolov reduction of a first order matrix differential
operator (i.e. a connection) with respect to a parabolic subgroup.
Here we show how the field content of such a theory can be obtained
starting from a solution of the Toda equations in the case of
systems defined on a higher genus \RS. We shall  confine
ourselves to the case of standard $A_{n-1}$--Toda equations.

Let us recall that
since a \RS\ has complex dimension 1, the $(0,1)$ part of any
connection $\nabla$ is automatically integrable, thus giving a
holomorphic structure to the complex vector bundle supporting it
\cite{AtBo82}. In this holomorphic frame one has
$\nabla^{\prime\prime}=\delb$. It will be convenient to refer to the
holomorphic bundle so obtained as the {\em holomorphic bundle defined
by (or associated to\/) $\nabla^{\prime\prime}$.\/}
Then, if the connection happens to be flat, its local
$(1,0)$--forms will be holomorphic in the holomorphic gauge. For a
complex vector bundle to admit a holomorphic connection is a
completely non trivial fact \cite{Ati57,Gun67}, since Weil's theorem states
that such a bundle must be a direct sum of indecomposable flat bundles.

We want now to construct the holomorphic
bundle (in the above sense) associated to the basic Higgs bundle
\rref{E:direct}, equipped with the connection
\begin{equation}\label{T-conn}
D^\prime = \del + \left(\begin{array}{ccccc}
			   \del\varphi_1&1& & & \\
                          &\del\varphi_2&1& & \\
			  & &\ddots & & \\
                          & & & &1\\
                          & & & &\del\varphi_n\end{array}\right)
\;
D^{\prime\prime} =\delb + \left(\begin{array}{ccccc}
		0& & & & \\
		\mbox{e}^{\varphi_1-\varphi_2}&0& & & \\
		&\mbox{e}^{\varphi_2-\varphi_3}&0& & \\
		& &\ddots &\ddots & \\
		& & &\mbox{e}^{\varphi_{n-1}-\varphi_n}&0\end{array}\right)
\end{equation}
(here $\sum_{i=1}^{n}\varphi_i=0$),
namely we want to prove the following
\begin{theorem}
The holomorphic vector bundle $V$ defined by the flat Toda connection
$D=D_{\! H}+\th +\th^*$ {\em is\/} the vector bundle of
$(n-1)$--jets of
sections of $K^{-\frac{n-1}{2}}$. The holomorphic connection $\nabla$,
which is the image of the Toda connection $D$ has the standard
W (or Drinfel'd--Sokolov) form:
\begin{equation}
\nabla^\prime = \del + \left(\begin{array}{cccccc}
			0&1& &         & & \\
              		 &0&1&         & & \\
             		 & & &\vdots   & & \\
             		 & & &         &0&1 \\
            	     w_n & w_{n-1}& &\cdots &w_2 &0
\end{array}\right)\, ,
\quad
\nabla^{\prime\prime}=\delb
\label{W:connection}
\end{equation}
with $\delb w_i=0$, $i=2,\dots n$.
\label{Toda-W}
\end{theorem}

\begin{rem}
We wish to stress the following point. The vector bundle $E$ is a
holomorphic bundle equipped with the $C^\infty$ connection $D$. Its
holomorphic structure is given simply by $D_{\!
H}^{\prime\prime}=\delb$. On the other hand $V$ is the holomorphic
vector bundle defined by the holomorphic structure $D^{\prime\prime}$.
Thus the two vector bundles $E$ and $V$ are {\em holomorphically
distinct,\/} although they are smoothly (i.e. $C^\infty$) equivalent.
\end{rem}
\begin{rem}
We  point out that the previous theorem explicitly constructs the
holomorphic vector bundle defined by $D''$, where $D$ is the Toda
connection, and identifies it with a concrete one (a holomorphic jet
bundle).
\end{rem}
The proof of theorem \ref{Toda-W} will be divided into steps,
or propositions, which are also of independent interest. More in details,
we want to show that the vector bundle $E$, associated in a suitable
covering $\{ {\cal U}_\alpha\}$ of $\Sigma$
by the $SL(n,{\Bbb C})$--cocycle
\beq
{\cal E}_{\al\be}=\left(\begin{array}{ccccc}
k_{\al\be}^{{n-1}\over 2}& & & & \\
&  k_{\al\be}^{{{n-1}\over 2}-1} & & & \\
& &\ddots & & \\
& & &    k_{\al\be}^{-{{n-1}\over 2}+1}& \\
& & & &  k_{\al\be}^{-{{n-1}\over 2}}\end{array}\right)
\eeq
equipped with the connection $D$ is $C^\infty$--equivalent
to the bundle $V$
of $(n-1)$--jets of sections of $K^{-{{n-1}\over 2}}$,
equipped with the connection
$\nabla$. We recall that the transition functions
${\cal V}_{\al\be}$ of $V$
can be gotten by expanding the relation
$\del^l_\al\xi_\al=( k_{\al\be}^{-1}
\del_\be)^l(k_{\al\be}^{{n-1}\over 2}\xi_\be)$,
$\xi_\al$ being a local section\footnote{For
$\Delta$ integer or half--integer we have
a well defined power (provided we make the choice of
a point of order $2$ in the Jacobian of $\Sigma$
if $\Delta$ is strictly half integer) $K^\Delta$ and that a (meromorphic or
$C^\infty$) section of $K^\Delta$ can be identified with a collection
of functions $\sigma_\al$ satisfying $
\sigma_\al = k_{\al\be}^{-\Delta}\sigma_\be$ in each overlapping
$U_\al\cap U_\be$.} of $K^{  -{{n-1}\over 2}}$,
and $k_{\al\be}=\deriv{z_\al}{z_\be}$.

We first discuss the transformation of $D$ into $\nabla$.
The following proposition is well--known, see
\cite{LeSa,bab89,GeO'RRaSa} and  it is implicitly contained in the
calculations of \cite{ABBP}. However, for the reader's convenience and
to avoid unpleasant gaps in the arguments proving Theorem
\ref{Toda-W},
we feel it worth stating it here and sketching its proof.
It is completely local in character.
\begin{prop}
There is a sequence $G_k$, $k=1,\dots ,n-1$, of gauge transformations,
taking their values in the lower nilpotent part of $SL(n,\CC )$, such
that the connection $D$ is mapped by $G=\prod_{k=1}^{n-1}\, G_k$ into
$\nabla$.
\label{local-equiv}
\end{prop}
\begin{pf}
It is a straightforward albeit long calculation, so that we only
illustrate the strategy and the first step. The idea is to show that
a column at a time can be operated on using (lower) nilpotent
abelian subalgebras of $sl(n,\CC )$. If $E_{i\, j}$ is the standard
matrix with $1$ in the $i\, j$ entry and zero elsewhere, we put
\begin{equation}
G = \prod_{k=1}^{n-1}\, G_k
\end{equation}
where
\begin{eqnarray*}
G_k &=& \exp \sum_{j=k+1}^n\, g_j^{(k)}E_{j\,k}\\
&=& \prod_{j=k+1}^n\, \exp {g_j^{(k)}E_{j\,k}}
\end{eqnarray*}
and recursively determine the $g_j^{(k)}$'s.

{}For instance, at step $\# 1$ we have to consider
$G_1=\exp\sum_{j=2}^n g_j^{(1)}E_{j\, 1}$
\begin{eqnarray*}
\left(D^{\prime\prime}\right)^{G_1} &=& (\delb g_2^{(1)}+
\mbox{e}^{\al_1(\Phi)})E_{2\, 1}\\
&+& \sum_{i=2}^{n-1}\mbox{e}^{\al_i(\Phi)}E_{i+1\, i}+\sum_{i=2}^{n-1}
(\delb g_{i+1}^{(1)}+g_{i}^{(1)}\mbox{e}^{\al_i(\Phi)}
)E_{i+1\, 1}
\end{eqnarray*}
so that the first column reads
\beq
\begin{array}{c}
\delb g_2^{(1)}+\mbox{e}^{
\al_1(\Phi)}\\ \cdot\\ \cdot\\
\delb g_{n}^{(1)}+g_{n-1}^{(1)}\mbox{e}^{\al_{n-1}(\Phi)}\end{array}
\eeq
As for the $(1,0)$ part of the connection we have
\begin{eqnarray*}
\left(D^{\prime}\right)^{G_1} &=& \sum_{i=1}^{n}\del\varphi_i E_{i\, i}+
g_2^{(1)} E_{1\, 1}+{\cal E}_+ +\\
&+& \sum_{j=1}^{n-2}\left(\del g_{j+1}^{(1)}+g_{j+1}^{(1)}(\del\varphi_{j+1}-
\del\varphi_1)-g_2^{(1)}g_{j+1}^{(1)}+g_{j+2}^{(1)}\right) E_{j+1\, 1}+\\
&+& \left(\del g_{n}^{(1)}+g_{n}^{(1)}(\del\varphi_{n}-
\del\varphi_1)-g_2^{(1)}g_{n}^{(1)}\right) E_{n\, 1}
\end{eqnarray*}
We can at once infer that $g_2^{(1)}=-\del\varphi_1$ and set to zero all the
coefficients of $E_{j+1\, 1},\> j=1,\ldots,n-2$ solving for $g_{j}^{(1)},
\> j=3,\ldots,n$ together with the first column of
$\left(D^{\prime\prime}\right)^{G_1}$, since, as it was proven
in~\cite{ABBP}, the Toda equations appear as the integrability
condition for such a system.

Thus the first step essentially boils down to produce the ``right''
first column of the connection matrices. One can repeat {\em verbatim}
the arguments above for the remaining columns. The consistency of the
procedure relies in the fact that at step $\# k$ we kill all the
elements of column $k$ in the $(0,1)$--part, we create the term in
the last row in the $(1,0)$ part while killing all others (except the
one in row $k-1$). It is not difficult to realize that such a
configuration is left invariant by the subsequent gauge transformations
$G_{k^\prime}$ for $k^\prime > k$.
\end{pf}

\noindent
The same procedure can be used to prove that the local gauge
transformations $G_i$ provide a $C^\infty$ cochain that sends the
cocycle ${\cal E}_{\al\,\be}$ into the cocycle ${\cal V}_{\al\,\be}$.
It is interesting to notice that, in our framework, the derivatives
$\del\varphi_i$ of the Toda fields appear as (minus)
the coefficients
$g_{i+1}^{(i)}$ of the negative simple roots in the
trivializing cochain. It is known that~\cite{bab89} the Toda equations
globalize on a higher genus \RS\ if the following non homogeneous gluing
law holds
\beq
\phi_{\al}^i=\phi_\be^i+{1\over 2}\sum_{l=0}^{i-1}(n-2 l -1)
\log|k_{\al\,\be}|^2\eeq
Indeed, the transformation law between ${\cal E}_{\al\,\be}$ and
${\cal V}_{\al\,\be}$ reproduces the correct factors for the local
fields $g_{\al\, i+1}^{(i)}$.

We now identify the holomorphic vector bundle defined by the Toda
connection. This is accounted for by the following
\begin{prop}
Let $V_n$ be a rank $n$ flat vector bundle admitting  a filtration
\begin{equation}
V_1^{(n)}\subset V_2^{(n)}\subset\cdots\subset V_n^{(n)}
\end{equation}
such that
$
V_{r+1}^{(n)}/V_r^{(n)}\simeq K^{({{n-1}\over 2}-r)}\; r=1,\ldots,n-1
.$
If $g(\Sigma)\ge 2$, then $V_n$ is the $(n-1)^{\mbox{th}}$--jet
extension of $K^{({{n-1}\over 2})}$.
\end{prop}

\begin{pf}

Let us consider the
sequence of quotients $\{K^{{{n-1}\over 2}},
K^{{{n-1}\over 2}-1},\ldots,K^{-{{(n-1)}\over 2}}\}$.
The last one gives
the sequence
\begin{equation}
0\to V_{n-1}^{(n)}\to V_n^{(n)}\to K^{-{{(n-1)}\over 2}}\to 0
\end{equation}
giving $[V_n]\in H^1(K^{{{n-1}\over 2}}
\otimes V_{n-1}^{(n)})$, where
$[E]$ denotes the equivalence class of $E$.
Tensoring with $K^{{{n-1}\over 2}}$ the sequences
\begin{equation}
0\to V_r^{(n)}\to V_{r+1}^{(n)}\to K^{{{n-1}\over 2}-r}\to 0,
\quad r=1,\ldots
,r-2
\end{equation}
and passing to the long cohomology sequences, we get the segments
\begin{equation}\label{como1}
H^1(V_r^{(n)}\otimes K^{{n-1}\over 2})\to H^1(V_{r+1}^{(n)}
\otimes K^{{n-1}\over 2})\to
H^1(K^{n-r-1})\to 0
\end{equation}
But $V_1^{(n)}=K^{{n-1}\over 2}$ so that (\ref{como1})
give $H^1(V_r^{(n)}\otimes K^{{n-1}\over 2})=0$ for $r=1,\ldots,n-2$,
and for $r=n-1$ we get the desired isomorphism
\begin{equation}
H^1(K^{{{n-1}\over 2}}\otimes V_{n-1}^{(n)})\simeq H^1(K)\simeq {\Bbb C}
\end{equation}
The non--triviality of the extension follows from
the fact that $V_2^{(n)}\subset V_{n}^{(n)}$ is
$K^{{{n-2}\over 2}}\otimes V_{2}^{(2)}$ and that the cocycle
defining the 1--jet extension of the spin bundle
\begin{equation}
0\to  K^{{{1}\over 2}}\to V_{2}^{(2)}\to K^{-{{1}\over 2}}\to
0\end{equation}
is one half of the first Chern class of $\Sigma.$

By Weil's theorem,
$V_n$ cannot be decomposable
into the direct sum of the powers of $K$ appearing in the diagonals,
and~\cite{NaRa69} any two non--trivial extensions of
$F_1$ by $F_2$ lying in the same ray in $H^1(\mbox{Hom}(F_1,F_2))$
give rise to isomorphic vector bundles. Thus $V_n$ is seen to be
isomorphic with the $(n-1)$--jet extension of $K^\frac{n-1}{2}$ once
one notices that the latter has the same sequence of quotients as in
the statement of the proposition.
\end{pf}
\subsection*{Example: the sl(3) case}

Let us examine in some details the $A_2$ ({\em alias}
$\frak{sl}(3,{\Bbb C})$) case in order to clarify how the picture
outlined above works.

The transition functions for the 2--jet bundle of
$K^{-1}$ (which is the case at hand) are given by the relations
\beq\label{holococ}
\left(\begin{array}{c}
\sigma_\al\\ \del_\al\sigma_\al\\ \del_\al^2\sigma_\al\end{array}
\right)
=\left(\begin{array}{ccc}
k_{\al\be}				 &0			&0\\
\del_\be \log k_{\al\be}		 &1			&0\\
k_{\al\be}^{-2}\del_\be^2 \log k_{\al\be}&k_{\al\be}^{-1}\del_\be
\log k_{\al\be}
&k_{\al\be}^{-1}\end{array}
\right)\,
\left(\begin{array}{c}\sigma_\be\\
\del_\be\sigma_\be\\ \del_\be^2\sigma_\be\end{array}\right)
\eeq
The smooth isomorphism between $V=J^2(K^{-1})$ and
$E = K^{-1}\oplus\CC\oplus K$
whose transition functions
${\cal E}_{\al\be}$ are the
diagonal part of eq.~(\ref{holococ}) is accomplished by
a {\em smooth} $SL(3,{\Bbb C})$-- valued $0$--cochain $G_\al$ which
we seek
in the factorized form (see the proof of prop. \ref{local-equiv})
\[
G_\al= G^{(1)}_\al G^{(2)}_\al
\]
with
\[
G^{(1)}_\al = \left(\begin{array}{ccc}
1			 &0			&0\\
h_\al		 	&1			&0\\
f_\al			&0			&1\end{array}
\right)\qquad
G^{(2)}_\al = \left(\begin{array}{ccc}
1			 &0			&0\\
0		 	&1			&0\\
0			&g_\al			&1\end{array}
\right)
\]

We will use the following standard overparametrization of the Toda field
by means of the triple $[\varphi_1,\varphi_2,\varphi_3]$ related
to the fields $\phi_1,\phi_2$ by $\phi_1=\varphi_1-\varphi_2,\> \phi_2=
\varphi_2-\varphi_3$.
Let us consider the Toda connection having the form
\begin{equation}
A^\prime=\left(\begin{array}{ccc}
\del \varphi_1 & 1 &0\\
0 &\del \varphi_2 &1 \\
0 & 0 &\del \varphi_1\end{array}\right)\qquad
A^{\prime\prime}=\left(\begin{array}{ccc}
0&0&0\\
\mbox{e}^{\varphi_1-\varphi_2}&0&0\\
0&\mbox{e}^{\varphi_2-\varphi_3}&0
\end{array}\right)\label{TC3}
\end{equation}
Under the transformation by $G^{(1)}_\al$ the cocycle~(\ref{holococ}) is
sent into one of the form
\[
\left(\begin{array}{ccc}
k_{\al\be}&0			&0\\
0	 &1			&0\\
0	 &k_{\al\be}^{-1}\del_\be\log k_{\al\be}&k_{\al\be}^{-1}
\end{array}\right)
\]
provided we have in the overlappings
\[
h_\al=k_{\al\be}^{-1} h_\be-k_{\al\be}^{-1}\del_\be\log
k_{\al\be}
\]
and
\[
f_\al=k_{\al\be}^{-2}f_\be+k_{\al\be}^{-1} h_\be\del_\be\log
k_{\al\be}-k_{\al\be}^{-3}\del_\be^2\log k_{\al\be}
\]
It is then easy to see that the reduction of the cocyle~(\ref{holococ}) to
its diagonal part is accomplished by the transformations
$G^{(2)}_\al$ provided $g_\al=k_{\al\be}^{-1} g_\be-k_{\al\be}^{-1}
\del_\be\log k_{\al\be}.$

More interesting is the transformation of the connection~(\ref{TC3})
which we display below dropping the indices referring to the coordinate
patches:
\[
(A^{\prime})^{G^{(1)}}=\left(\begin{array}{ccc}
\del\varphi_1+h&1&0\\
\del h +h(\del\varphi_2-\del\varphi_1)+f-h^2& \del\varphi_2-h &1\\
\del f +f(\del\varphi_3-\del\varphi_1)-h f& -f &0
\end{array}\right)\]\[
(A^{\prime\prime})^{G^{(1)}}=\left(\begin{array}{ccc}
0 &0 &0\\
\mbox{e}^{\varphi_1-\varphi_2}+\delb h&0&0\\
h \mbox{e}^{\varphi_2-\varphi_3}+\delb f&\mbox{e}^{\varphi_2-\varphi_3}
&0\end{array}\right)
\]
This means that we have to solve for the equations
\begin{eqnarray*}
\del\varphi_1+h=0& &\del h +h(\del\varphi_2-\del\varphi_1)+f-h^2=0\\
\mbox{e}^{\varphi_1-\varphi_2}+\delb h=0& &h
\mbox{e}^{\varphi_2-\varphi_3}+\delb f=0
\end{eqnarray*}
which gives $h=-\del\varphi_1$ and $f=\del^2\varphi_1+\del\varphi_1
\del\varphi_2,$ the other two equations being identically true on the
solutions of the Toda equations.
The action of the subsequent gauge transformation
$G^{(2)}$ gives $g=\del\varphi_3$ and sends the Toda connection into the
form
\beq
(A^{\prime})^{G}=\left(\begin{array}{ccc}
0&1&0\\
0&0&1\\
w_3&w_2&0\end{array}\right)\, ,\qquad
(A^{\prime\prime})^{G}=0
\eeq
where $w_2$ is the usual energy momentum tensor of the
${\goth sl}_3$ Toda theory,
\beq
w_2=(\del\phi_1)^2+(\del\phi_2)^2-[\del^2\phi_1+\del^2\phi_2+\del\phi_1\del
\phi_2]
\eeq
and
\beq w_3=\del w_2+(\del\phi_1)^2\del\phi_2
-\del\phi_1(\del\phi_2)^2+2\del\phi_2\del^2\phi_2-\del^3\phi_2
\eeq
is the other generator of the $W_3$--algebra.

\section{Embeddings and $W$--Geometry}
The purpose of this section is to discuss more thoroughly the
geometrical features of (standard) Toda Field Theory. In particular, we
shall discuss the embeddings of a \RS\ determined by the
classifying map resulting from the self--duality equations, see
section \ref{higgs}. The crucial properties for this analysis are
the natural filtration of the Higgs bundle, and the fact that, at least
in the standard Toda case, there is an additional real structure
preserved by the Toda connection.
As in section \ref{W-section}, our discussion will be
confined to the $A_{n}$ case.

\subsection{The additional real structure}
\label{add:real}
Consider the basic Higgs system given by \rref{theta} and
\rref{E:direct} together with the harmonic metric $H$. By the general
discussion in section \ref{higgs} we
know that the structure group of $E$ as a harmonic bundle reduces to
$SL(n,\RR)$. We now show that there is another real structure
compatible with this one.
Let $A:E\rightarrow E$ be the endomorphism equal to
$(-1)^r$ on each factor $K^{-\frac{n-1}{2}+r}$. With it, we construct
an indefinite hermitean form $<\cdot ,\cdot >$ over $E$, namely
\begin{equation}
<u,v>={(A\, u,v)}_H\, ,\qquad u,v\in E
\label{<>}
\end{equation}
A straightforward calculation proves
\begin{lem}
\label{flat-indef-form}
The hermitean form \rref{<>} is flat with respect to the Toda
connection $D=D_{\scriptstyle H}+\th +\th^*$, that is we have
\[
\mbox{d}\, <u,v>\, =\, <D\, u,v> + <u,D\, v>\,\qquad u,v\in E
\]
\end{lem}

This implies, of course, that a reduction of the structure group from
$SL(n,\CC)$ to $SU(p,q)$, where $p=[n/2],\, q=n-p$, takes place. More
precisely, what we actually mean by ``$SU(p,q)$'' is the group
corresponding to the fixed point set in ${\frak g}^\CC =A_{n-1}$ of the
conjugation $\nu$ given by
\begin{equation}
\nu (\xi )= - I\,\rho (\xi )\, I\, ,\qquad \xi\in{\frak g}^\CC
\label{nu}
\end{equation}
where in this case $\rho$ is simply minus the hermitean conjugate and
$I$ is the diagonal matrix
\begin{equation}
I=
\left(\begin{array}{cccc}
1& & & \\
 &-1& & \\
 &  &\ddots& \\
 &  &      &\pm 1
\end{array}\right)
\label{I}
\end{equation}
(the sign in the last element being determined according to the parity of
$n$). It is obvious that this is the standard form for $SU(p,q)$ up to
a coordinate reshuffling. It is also easy to see that the conjugations
$\tau$ defined in section \ref{higgs} (eq. \rref{tau}) and $\nu$
commute, so that (the Lie algebra of) the structure group of the
harmonic bundle corresponding to the Toda equations is in fact given
by the intersection of the fixed point sets of $\tau$ and $\nu$.
Let us call $G$ the
real structure group so obtained. We further define $K$ to be its
maximal compact subgroup.

By the results about harmonic bundles quoted in section \ref{higgs}, we
thus obtain a harmonic map
\[
f_H:\widetilde{\Sig}\longrightarrow G/K
\]
where $\widetilde{\Sig}$ is the universal cover of $\Sig$, or, in
other terms, a map
\[
f_H:\Sig\longrightarrow\Ga\backslash G/K
\]
where the discrete subgroup $\Ga \subset G$ is the image of
$\pi_1(\Sig )$ through the holonomy. By a little abuse of language, we
use the same symbol for both.
For the sake of convenience, let us denote $\widetilde{N} =
G/K$ and $N = \Ga\backslash G/K$. The stability properties of the
Higgs bundle we are looking at \cite{Hit92} imply the action of $\Ga$
on $\widetilde N$ to be properly discontinuos, so that $N$ is a manifold.

Thus we can interpret the Toda field equations
as the equations characterizing the embedding of the \RS\ into a some
homogeneous
manifold $N$ through a harmonic map $f_H$. We can actually refine
this, that is  starting from the map $f_H$ or
-- what is the same -- from the harmonic bundle we can construct an
embedding $F:\Sig\rightarrow\DD$ into a {\em complex\/} manifold $\DD$.
This requires analyzing more extensively the structure of
the bundle we associated to the Toda equations.

\subsection{Toda systems and Variations of Hodge structures}
Upon rewriting our rank--$n$ basic bundle \rref{E:direct}
as \cite{DeBGoe}
\begin{equation}
E=\bigoplus_{r+s=n-1}\,E^{r,s}\, ,\qquad E^{r,s}=K^{-\frac{n-1}{2}+r}
\label{hodge:1}
\end{equation}
the Higgs field $\th$ appearing in the Toda connection, eq.
\rref{theta}, has  the property
\begin{equation}
\th :E^{r,s}\longrightarrow E^{r-1,s+1}\otimes K
\label{hodge:2}
\end{equation}
and the factors are orthogonal with respect ot both the metric $H$ and
the indefinite hermitean form $<\cdot,\cdot >$.
As a consequence, the complete connection $D=D_{\! H}+\th + \th^*$
satisfies the following {\em Griffiths transversality condition\/}
\begin{equation}
D:E^{r,s}\longrightarrow
A^{1,0}(E^{r-1,s+1})\oplus A^{1,0}(E^{r,s})\oplus
A^{0,1}(E^{r,s})\oplus A^{0,1}(E^{r+1,s-1})
\label{hodge:3}
\end{equation}
where by $A^\bullet (E^{r,s})$ we mean $C^\infty$ sections. It is
useful for later purposes to rewrite \rref{hodge:3} in the following
form. Consider the filtration
\[
E\equiv F^0\supset F^1\supset\cdots\supset F^{n-1}\supset F^{n}\equiv
\{ 0\}
\]
where
\beq
\label{filter}
F^q=\bigoplus_{r=q}^{n-1}\,
K^{-\frac{n-1}{2}+r}=\;\bigoplus_{r=q}^{n-1}\, E^{\, r,s}
\eeq
Then the transversality condition can be restated as
\begin{eqnarray}
& &D^\prime :F^q\longrightarrow A^{1,0}(F^{q-1})\nonumber\\
& &D^{\prime\prime}:F^q\longrightarrow A^{0,1}(F^q)
\label{hodge:4}
\end{eqnarray}
Notice that $\mbox{rk}\, F^q=n-q$ and that these are the
subbundles\footnote{Up to a reshuffling of indices}
corresponding to the filtration of $V$ by the $V_q^{(n)}$'s
appeared in section \ref{W-section}. In the purely
holomorphic picture \rref{hodge:4} reads
\begin{equation}
\nabla^\prime : V^{(n)}_q\longrightarrow \Omega^1(V^{(n)}_{q-1})
\label{hodge:5}
\end{equation}
where $\Omega^1(\cdot)$ denotes the space of
holomorphic differentials.

According to Simpson, a harmonic bundle $E=\oplus_{r+s=w}\,E^{r,s}$
whose factors are orthogonal with respect to an indefinite hermitean
form $<\cdot ,\cdot >$, satisfying  any one of
\rref{hodge:2}--\rref{hodge:5} defines a {\em complex variation of Hodge
structure\/} \cite{Simp88,Simp92,Gri70}\footnote{The main
difference from Griffiths' original
definition is that in Simpson's one the existence of the integral
lattice is left out. We shall stick to this one.} of weight $w$.

Thus the Higgs bundle associated to the Toda equations displays the
formal properties of a \VHS\ of weight $w=n-1$, whose ``Hodge Bundles''
$E^{\, r,s}=K^{-\frac{n-1}{2}+r}\cong F^r/F^{r+1}$ are in fact line
bundles. We shall use the machinery of \VHSs\ to prove
\begin{theorem}
The Toda equations determine a holomorphic embedding
\[
F_H:\Sig\longrightarrow \Ga\backslash\DD
\]
where $\Ga$ is the monodromy group, $\DD\cong G/K_0$ a Griffiths period
domain, $G$ is the structure group
defined in \S 5.1
and $K_0\subset K\subset G$ a
(compact) subgroup. The map $F_H$ {\em is} the metric $H$ seen as a
section of a flat bundle over $\Sigma$ with typical fibre $G/K_0$ and
its differential is given by the Higgs field $\th$.
\label{embeddings}
\end{theorem}
For the proof we need to recall some basic properties of
Griffiths period domains (or classifying spaces).

\subsubsection*{A brief tour through period domains}

Let us denote by ${\bf E}$ a complex
vector space equipped with
\begin{itemize}
\item a conjugation
$\cdot^\sig :{\bf E}\rightarrow {\bf E}$
\item a bilinear form $Q:{\bf E}\times{\bf E} \rightarrow\CC$ such that:
\begin{enumerate}
\item $Q(v,u)=(-1)^{w}Q(u,v)$, $u,v\in {\bf E}$,
\item it is ``real'' with respect to the conjugation of ${\bf E}$, namely
$\overline{Q(u,v)}=Q(u^\sig ,v^\sig )$,  $u,v\in {\bf E}$.
\end{enumerate}
\end{itemize}
A period domain $\DD$ is the set of all
weight $w$ Hodge structures on ${\bf E}$,
namely the set of all decompositions ${\bf E}=\oplus_{r+s=w}\,
{\bf E}^{\, r,s}$
satisfying
\[
\begin{array}{cl}
Q({\bf E}^{\, r,s},{\bf E}^{\, r^\prime ,s^\prime })=0 &
\mbox{unless $r^\prime =
s$ and $s^\prime = r$}\\
i^{r-s}Q(u,u^\sig )>0 & \mbox{for any $u\in {\bf E}^{\, r,s}$}
\end{array}
\]
If the weight $w$ is $n-1$, which is the case we will be dealing with,
then all
the factors in a given Hodge structure are actually lines, but
this is
not quite so in general. It is useful to define the same object
in terms of filtrations. To this end, define the operator
$C:{\bf E}\rightarrow {\bf E}$ by $C\mid_{{\bf E}^{\, r,s}}=i^{r-s}$.
Then $\DD$ is defined to be the set of all (descending) filtrations
$\{ {\bf F}^q\}$  in ${\bf E}$ such that
\begin{equation}
\begin{array}{c}
Q({\bf F}^q, {\bf F}^{w-q+1})=0\\
Q(C u, u^\sig )>0
\end{array}
\label{Riem-Gri}
\end{equation}
The link between the two definitions lies in the decomposition
${\bf F}^q=\oplus_{r=q}^{w}\,{\bf E}^{\, r,s}$. Dropping the second
condition in either one of the two definitions we just gave, yields
the {\em compact dual\/} $\check{\DD}$ of $\DD$. It is an algebraic
subvariety of a flag manifold, and hence of a product of Grassmannians
\cite{Gri84}. This is clear from the second definition. This implies
for $T_{\check{\DD}}$, the holomorphic tangent bundle of $\check{\DD}$, that
\begin{eqnarray*}
T_{\check{\DD}} &\subset &
\bigoplus_{q=1}^w\,\mbox{Hom}({\bf F}^q,{\bf E}/{\bf F}^q)\\
&=& \bigoplus_{q=1}^w\,\bigoplus_{r=1}^q\,\mbox{Hom}({\bf E}^{\,
q,w-q},{\bf E}^{\, q-r,w-q+r})
\end{eqnarray*}
The group
$G^\CC = \mbox{Aut}({\bf E},Q)$ acts transitively on it \cite{Gri70}, thus
$\check{\DD}$ is in fact a complex {\em manifold.\/} The period domain
$\DD$ lies inside it as an open subset, and therefore as a complex
submanifold. It is the open orbit through the origin of
$\check{\DD}$ of the real form of
$G_\CC$ with respect to the given conjugation in ${\bf E}$.

\subsubsection*{Proof of Theorem \protect\ref{embeddings}}
We start with
\begin{lem} The group $G$ introduced in subsection \ref{add:real} is
the real group acting on the classifying space of weight $n-1$ Hodge
structures on a vector space of dimension $n$.
\label{bip}
\end{lem}
\begin{pf}
The choice of a basis $e_1,\dots ,e_n$ in a complex vector field ${\bf
E}$ of dimension $n$ allows to define the
decomposition
\[
{\bf E}=\bigoplus_{r+s=n-1}\, {\bf E}^{\, r,s}\, ,\qquad {\bf E}^{\,
r,s}=\CC \{e_{r+1}\}
\]
and an indefinite hermitean form through
\[
<e_i,e_j>=\delta_{ij}(-1)^{i+1}
\]
Then the form $<\cdot ,\cdot >$ is the one represented by
the matrix $I$
of \rref{I}.
We can moreover consider the conjugation map
\begin{eqnarray*}
\cdot^\sig :&{\bf E}&\longrightarrow {\bf E}\\
&u&\mapsto u^\sig = S\bar u
\end{eqnarray*}
where $S$ is the matrix \rref{S}. Notice that with this definition we
have $({\bf E}^{\, r,s})^\sig ={\bf E}^{\, s,r}$.
Thus the above decomposition is
a (reference) Hodge structure in ${\bf E}$ of weight $n-1$
\cite{Gri70,Gri84}.  Then we define the bilinear form $Q$ by
\[
Q(u,v)=i^{n-1}<u,v^\sig >\, ,\qquad u,v\in {\bf E}
\]
It is easy to check that it has the properties listed before in the
r\'esum{\'e} of period domains.
We finally introduce the complex Lie group $G_\CC = SO(Q,\CC )\equiv
\mbox{Aut}({\bf E},Q)$ of those complex automorphisms of ${\bf E}$
which preserve $Q$.

Recall now that $G$ has been defined as the real group arising as
intersection of the fixed point sets of the conjugations $\tau$ and
$\nu$ in \rref{tau}, \rref{nu}. Therefore it preserves the
hermitean form $<\cdot ,\cdot >$.
It is then obvious that $G$ coincides with the real subgroup of
$G_\CC$ defined by
the conjugation $\cdot^\sig$, as we clearly have $(g\, u)^\sig = \tau
(g)\, u^\sig$, for $u\in {\bf E}$ and $g\in G_\CC$.

The statement now
follows from the fact that the period domain $\DD$ is the
quotient $G/K_0$ \cite{Gri70,Gri84} where $K_0=G\cap B$ and $B\subset
G_\CC$ is the stabilizer of the reference Hodge structure.
\end{pf}
\begin{rem} The stabilizer group $K_0$ is in general {\em strictly\/}
contained in the maximal compact subgroup $K$ of $G$.
\end{rem}
Let us now come back to the Higgs bundle $E$ equipped
with the filtration $\{ F^q\}$
defined in equation~\rref{filter}.

To construct the mapping $F_H:\Sig\rightarrow\Ga\backslash \DD$,
choose a basepoint $x_0$ on $\Sig$. Then we look at the fibre
$E_{x_0}$ as the fixed vector space ${\bf E}$. Notice that the
conjugation in the
proof of the preceding lemma agrees with the one in $E$ constructed in
section \ref{higgs}.
Thus we repeat the constructions in the proof of lemma \ref{bip} and
get the hermitean form $<\cdot ,\cdot
>_{x_0}$ and the required bilinear form $Q$ as well. Since the
connection $D$ is flat, we can use the isomorphism of any fibre $E_x$
with $E_{x_0}$ to induce a filtration on $E_{x_0}$ (which will be the
image of the filtration $\{ F^q_x\}$ on $E_x$).
The hermitean form $<\cdot,\cdot>$ on $E$ is {\em flat\/}
under $D$, thus the orthogonality
properties of the subspaces in the fibre $E_x$ translate into the
bilinear relations for the induced filtration on $E_{x_0}$. Therefore the
reference Hodge structure in $E_x$ defines a new Hodge structure in
$E_{x_0}$ and we obtain a local map from $\Sigma$ to $\DD$.
The entire
construction is of course defined up to the action of the monodromy
group.
\begin{lem}
The mapping $F_H:\Sig\rightarrow\Ga\backslash \DD$ so defined is
holomorphic.
\end{lem}
\begin{pf}
The statement is local. By construction, the differential of $F_H$ is
defined by the flat connection $D$ itself. From the inclusion
$\DD\subset\check{\DD}$ we have~\cite{Gri84}
\[
T_{\DD} \subset
\bigoplus_{q=1}^{n-1}\,\mbox{Hom}({\bf F}^q,{\bf E}/{\bf F}^q)
\]
and using this picture
for the tangent bundle, holomorphicity follows from the transversality
condition $D^{\prime\prime}:F^q\rightarrow A^{0,1}(F^q)$.
\end{pf}
\begin{rem}
According to the given description of $T_{\check{\DD}}$, the other half
of the transversality condition says that ${F_H}_* (T_\Sig )\subset
\bigoplus_{r=1}^{n-1}\,
\mbox{Hom}({\bf E}^{\, r,n-1-r},{\bf E}^{\, r-1,n-r})$. This is the
geometrical meaning of the {\em grading condition\/} imposed on the
Toda connection \cite{GeO'RRaSa,RaSa93}.
\end{rem}
Finally, we can conclude that the map $F_H$ is
essentially the metric
$H$, seen as a section of a bundle in homogeneous spaces,
 by applying the same argument we used in section \ref{higgs}. The
key point is to notice that the \VHS\  defines a reduction of the
structure group $G$ to its intersection with the stabilizer of the
reference Hodge structure (the group we called $K_0$ before)
\cite{Simp88}, and it is obvious that the section so obtained
coincides with the metric $H$. The theorem is
proved.\hfill$\square$
\begin{rem}
We notice that the targets of the holomorphic embeddings so
obtained {\em depend on the genus\/} of the Riemann Surface, see
Section 6 below. This is a
consequence of the holonomy representation of the fundamental group on
the Griffiths period domain $\DD$.
\end{rem}
The situation with our embeddings is the following. Using the
fibration $\DD\cong G/K_0\rightarrow G/K\equiv\widetilde{N}$
\cite{Gri70} we have the diagram
\[
\setlength{\unitlength}{0.0070in}%
\begin{picture}(171,106)(200,540)
\put(382,621){\vector( 0,-1){ 60}}
\put(228,558){\vector( 2, 1){132}}
\put(228,546){\vector( 1, 0){132}}
\put(205,540){\makebox(0,0)[lb]{\raisebox{0pt}[0pt][0pt]{$\Sigma$}}}
\put(368,628){\makebox(0,0)[lb]
{\raisebox{0pt}[0pt][0pt]{$\Gamma\backslash\DD$}}}
\put(371,540){\makebox(0,0)[lb]{\raisebox{0pt}[0pt][0pt]{$N$}}}
\put(290,520){\makebox(0,0)[lb]{\raisebox{0pt}[0pt][0pt]{$f_H$}}}
\put(290,610){\makebox(0,0)[lb]{\raisebox{0pt}[0pt][0pt]{$F_H$}}}
\end{picture}
\]
where the map $f_H$ is harmonic ($N$ is in general not complex) and
$F_H$ is holomorphic. This is an instance of a more general situation
in which harmonic maps are covered by \VHSs\ \cite{CaTo89}.

The complete classification of the homogeneous spaces for the $A_n$ case
has been performed by Griffiths~\cite{Gri70}, as
\[
\DD = \left\{
\begin{array}{ll}
SO(n+1,n,\RR)/U(1)^n & \mbox{for $A_{2n}$}\\
Sp(2 n , \RR)/U(1)^n & \mbox{for $A_{2n-1}$}
\end{array}\right.
\]
Our favourite example of $A_2$ can be easily worked out completely.
In this case we $Q$ is represented by the matrix
\beq
\left(\begin{array}{ccc}
& & -1\\
& 1 &\\
-1& & \end{array}\right)\eeq
Looking at the left half of the Hodge diamond,
the first bilinear relation~(\ref{Riem-Gri}) yields the divisor $2 X_0 X_2-
X_1^2=0$ in $\PP^2$, the rational normal curve given by the Veronese
embedding, while the second selects the complement of the real
circle $|z|^2=2$ in this rational curve, i.e. we have the disjoint union of
two copies of the Poincar\'e disk.

We shall conclude this section with a brief comment on the
Conformal Affine case. Most of the structure described
in this section does not carry over directly to this more general case. In
particular, lemma \ref{flat-indef-form} is easily seen to be false if
$\th$ is given by \rref{theta:aff}. Therefore the second real
structure cannot be defined by means of the endomorphism
I, which implies that we cannot follow the path of the standard Toda case
to define a
\VHS\ and we cannot use the holomorphic embedding into the Griffiths
period domain any longer.

A possible way out can be conceived along the following lines.
The real structure  described in section \ref{higgs} is not ruled
out by the deformation leading to the Conformal Affine Toda system
and
therefore there is still a map
\[
f_H:\widetilde{\Sigma}\longrightarrow SL(n,\RR )/SO(n)
\]
Being the target manifolds not complex, the map $f_H$ above
is not suitable as it stands to construct holomorphic embeddings.
Anyhow, since the map $f_H$ is
harmonic and $\Sigma$ has complex dimension $1$,
we will have
\beq
{\widetilde\nabla}^{\prime\prime}\del f_H =0
\label{harmo}
\eeq
where in this case $\del f_H$ is to be understood as a section
of the vector bundle
\[ T^*_\Sigma\otimes f_H^*(T^\CC_{SL(n,\RR)/SO(n)})\]
and $\widetilde\nabla$ is the tensor
product of the K{\"a}hler connection on $K\equiv
T^*_\Sigma$ and the pull--back of the
Riemannian one on $T^\CC_{SL(n,\RR)/SO(n)}$ respectively \cite{CaTo89}.
Thus $\del f_H$ is a holomorphic section of a certain complex vector
bundle over $\Sigma$. Since the map $f_H$ is determined by the metric
$H$, by equation \rref{alpha} we have that equation~(\ref{harmo}) is
the translation in this framework of $\delb\th =0$.

\section{Conclusions and comments}

In this paper we have analized the ``triangular''
 correspondence~\cite{GeO'RRaSa,Hit87}
\[
\mbox{Toda} \leftrightarrow W_n-\mbox{algebras} \leftrightarrow
\mbox{Higgs bundles}
\]
from the
point of view of the theory of hermitean holomorphic vector bundles on a
generic genus \RS\ $\Sigma$. Although the origin of such relationships can
be traced back
to the fact that both the Toda Field and the Hitchin's self--duality
equations are
dimensional reductions of a suitable four dimensional Self Dual Yang Mills
theory, we deemed it worthwhile
to work out some topics from the two--dimensional
viewpoint.

In particular we have proved that the assignement of a solution of
the $A_{n-1}$ Toda Field equations determines both in standard
and in the Conformal Affine case a harmonic Higgs bundle.
The metric is parameterized by the Toda fields, and the Higgs field
arises as the non metric part of the total flat connection.

The underlying holomorphic vector bundle $V$ is
uniquely fixed to be the
bundle of $(n-1)^{th}$--jets of sections of $K^{-(n-1)/2}$. $W_n$ fields
are naturally identified with the non trivial entries of the flat
{\em analytic} connection on V.
Actually, this is not an unexpected feature and it has
already been found out f.i.
in~\cite{BiGe89};
we would like however
to point out that
a very nice geometrical
significance of the Toda fields as the building blocks of the local
isomorphisms (in the $C^\infty$--category) between the two holomorphically
distinct bundles $E = \bigoplus_{r = 0}^{n-1} K^{-\frac{n-1}{2}+r}$ and
$V=J(K^{-\frac{n-1}{2}})$ is enlightened, together with some global
features
of the higher genus case which were perhaps a bit overlooked in the
literature.

The main point can be considered the discussion of how the
datum of a $A_{n-1}$--Toda Field on $\Sigma$ leads, in the standard
case, to the
realization of the
Riemann Surface as a base space for a Variation of Hodge Structure of
weight $n-1$ and rank $n$, and henceforth a holomorphic map from $\Sigma$
into a quotient of a Griffiths period domain $G/K_0$.

Since these results go in the direction of the so called geometry of
$W_n$--embeddings as put forward by Gervais, Saveliev and collaborators,
a couple of comments are in order.

First of all, in the paper~\cite{RaSa93}
the following picture is explained. The
starting point is a $C^\infty$--map from $\Sigma$ to a complex Lie
group $G$ which, under suitable instances (the ``grading condition'') lifts
a holomorphic map $\varphi_P :\Sigma \to G/P$, $P$ being a parabolic
subgroup of $G$. Considering those parabolic subgroups
$P_i, i=1,\ldots ,\mbox{rank}\, G,$ for which $G/P_i$ is the $i^{th}$
fundamental
homogeneous space for $G$, the associated maps
$\varphi_{P_i}$ define maps from $\Sigma$
to $\PP (V_i)$, the projectivisation of the  $i^{th}$ fundamental
representation of $G$.
Then it is shown  that the (generalized) Pl\"ucker relations
for the curvature of the pull--back on $\Sigma$
of the Fubini--Study metrics on  $\PP (V_i)$
on $\Sigma$ translate, when expressed through local K\"ahler potentials,
into the
Toda Field equations for a suitably chosen local representative of $
\varphi_{P_i}$.

Our starting point is different: we {\em start} from a solution of
the Toda Field equations and we {\em determine}
a holomorphic map from $\Sigma$ to a suitable locally homogeneous
space.
It follows that the target space
we obtain is only {\em locally} determined by the rank of the Cartan
subalgebra in which the Toda fields take values, since in the large
the monodromy action
of the fundamental group of $\Sigma$
on the Griffiths period domain must be factored out, thus yielding
a different global target space according to the genus $g(\Sigma)$.

Nonetheless, Pl\"ucker formulas are of local type,
so one should expect them to
arise also in our context. Indeed, one can see that they  can be
recovered by considering the natural embeddings of the algebraic
manifold $\check{\DD}$ into the product of Grassmannians~\cite{Gri84}
\[
G(h_1,n)\times G(h_2,n)\cdots \times G(h_{n-1},n)
\]
($h_r$ is the rank of $F^r$ in the Hodge filtration)
and pulling back to $\check{\DD}$ the determinant line bundles
associated to the tautological sequences
\[
0\longrightarrow S_{h_r}\longrightarrow {\CC}^n\longrightarrow
Q_{h_r}\longrightarrow 0
\]
over $G(h_r,n)$. Pl\"ucker coordinates for $G(h_r,n)$ are indeed
obtained by taking holomorphic sections of $\mbox{det}\, Q_{h_r}.$
It is to be borne in mind that $\check{\DD}$ is {\em
strictly\/} contained in the complete flag manifold for $SL(n,\CC )$:
the $A_{n-1}$--type Pl\"ucker formulas ensuing from the various
tautological sequences can be obtained by explicitly realizing the
embedding of $\check{\DD}$ in the product considered above.

The following, and final, remark is also partly motivated by the last
observation. We have shown that the Toda connection
$D$ is compatible with the two Lie algebra automorphisms \rref{tau}
and \rref{nu} which we rewrite here:
\bea
\tau (X) &=& S \bar{X} S\nonumber\\
\nu (X) &=& - I \bar{X^t} I\nonumber
\eea
Following \cite{FrRaSo93}, we
notice that $\hat{\tau}=\tau\nu = \nu\tau$ is the extension of the
automorphism of the A--type
Dynkin diagram which can be used to define the algebras $B_r$ and
$C_r$ as quotient of $A_{2r}$ and $A_{2r-1}$ respectively. Moreover,
since the
Griffiths
period domains in which $\widetilde\Sigma$ is immersed are quotients of
$SO(r+1,r)$ and $Sp(2r,\RR )$ respectively, we are lead to
conclude that the maps induced by
the metric might give insights into the geometry of the $W(B)$ and
$W(C)$--algebras obtained via folding procedure by the original
$A$--theory.

We hope to discuss more thoroughly those questions
in a future work.

\subsection*{Acknowledgements} One of us (E. A.) warmly
thanks L. Bonora
and J--L. Dupont for useful discussions.


\begin{thebibliography}{10}
\bibitem{ABBP}
E. Aldrovandi, L. Bonora, V. Bonservizi and R. Paunov, {\em Free Field
representation of Toda Field Theories.\/} \ijmp{9}{1994}{57--86}
\bibitem{AFGZ91}
H. Aratyn, L.A. Ferreira, J.F. Gomes, A.H Zimerman, {\em Kac--Moody
construction of Toda type field theories.\/}
\pl{254}{1991}{372}
\bibitem{Ati57}
M. F. Atiyah, {\em Complex analytic connections in fibre bundles.\/}
\tams{85}{1957}{185--207}
\bibitem{AtBo82}
M. F. Atiyah, R. Bott, {\em The Yang Mills Equation over
\RSs.\ } \newblock Phil. Trans. Roy. Soc. London Series A
{\bf 308} (1982), 524--615
\bibitem{bab89}
O. Babelon, {\em From Integrable to Conformal Field Theory.\/}
Preprint PAR LPTHE 89/29
\bibitem{BaBo90}
O. Babelon, L. Bonora, {\em Conformal affine $sl_2$ Toda field
theory.\/} \pl{244}{1990}{220--226}
\bibitem{BaBoScSu88}
F. A. Bais, P. Bouwknegt, K. Schoutens, M. Surridge, {\em
Extensions of the Virasoro algebra constructed from Kac--Moody algebras
using higher order Casimir invariants.\/} \np{304}{1988}{348}.
\bibitem{BiFoKo92}
A. Bilal, V. V. Fock, Yu. I. Kogan, {\em On the Origin of
$W$--algebras.\/}
\np{359}{1991}{635--672}
\bibitem{BiGe89}
A. Bilal, J--L. Gervais, {\em
Extended c=$\infty$ conformal systems
from classical Toda field theories.\/} \np{314}{1989}{646--686}
\bibitem{CaTo89}
J. A. Carlson, D. Toledo, {\em Harmonic mappings of K{\"a}hler
manifolds to locally symmetric spaces.\/} \ihes{69}{1989}{173--201}
\bibitem{CeVa91}
S. Cecotti, C. Vafa, {\em Topological -- antitopological
fusions.\/} \np{367}{1992}{359--461}.
\bibitem{Cor88}
K. Corlette, {\em Flat $G$--bundles with canonical metrics.\/}
J. Diff. Geom {\bf 28} (1988), 361--382
\bibitem{DeBGoe}
J. de Boer, J. Goeree, {\em Covariant $W$-gravity \& its moduli space
from gauge theory.\/} \lanl{9206098}
\bibitem{Don87}
S. K. Donaldson, {\em Twisted harmonic maps and self-duality
equations.\/}
\plms{55}{1987}{127--131}
\bibitem{EeSam64}
J. Eells, J. H. Sampson, {\em Harmonic mappings of Riemannian
manifolds.\/} \ajm{86}{1964}{109--160}
\bibitem{FaLu88}
V. A. Fateev, S. L. Lukyanov, {\em Additional symmetries and
exactly--soluble models in two--dimensional conformal field theory.\/}
Sov. Sci. Rev. A. Phys. {\bf 15} (1990), 1--124.
\bibitem{FrRaSo93}
L. Frappat, E. Ragoucy, P. Sorba {\em Folding the
$W$-algebras.\/}\np{404}{1993}{805}
\bibitem{Fr91}
E. V. Frenkel, {\em W--algebras and Langlands--Drinfel'd correspondences.\/}
Proceedings of the Carg\`ese Summer School {\em New Symmetry Principles
in QFT}. July 1991.
\bibitem{GeNeYY}
J.--L., Gervais, A. Neveu,
{\it The dual string spectrum in
Polyakov's quantization. I.\/} \np{199}{1982}{59};
{\it The dual string spectrum in
Polyakov's quantization. II.\/} \np{209}{1982}{125}.
\bibitem{GeMa92}
J.--L. Gervais, Y. Matsuo, {\em Classical $A_n$--W--Geometry.\/}
\cmp{152}{1993}{317--368}
\bibitem{GeO'RRaSa}
J.--L. Gervais, L. O'Raifeartaigh, A. V. Razumov, M. V. Saveliev, {\em
Gauge conditions for the constrained WZWN--Toda reductions.\/}
\lanl{9211088}
\bibitem{GeSa93}
J.--L. Gervais, M. V. Saveliev, {\em W--Geometry of the Toda
systems associated with non--exceptional simple Lie algebras.\/}
\lanl{9312040}
\bibitem{Gri70}
P. Griffiths, {\em Period of integrals on algebraic manifolds\/}~III.
\ihes{38}{1970}{125--180}
\bibitem{Gri84}
P. Griffiths et al., {\em Topics in trascendental algebraic
geometry.\/} Princeton Univ. Press, Princeton, NJ, 1984.
\bibitem{Gun67} R.C.Gunning,
{\em Lectures on Vector bundles over Riemann Surfaces.\/}
Princeton. Univ. Press., Princeton 1967.
\bibitem{Hit87}
N. J. Hitchin, {\em Self duality Equations on a Riemann
Surface.\/}
Proc. London Math. Soc. {\bf 55}, (1987), 59 -- 126.
\bibitem{Hit87b} N. J. Hitchin, {\em Stable bundles and integrable systems.\/}
Duke Math. Journal {\bf 54}, (1987), 91--114.
\bibitem{Hit92}
N. J. Hitchin, {\em Lie Groups and Teichm\"uller Theory.\/}
Topology {\bf 31}, (1992), 451 -- 487.
\bibitem{Kos59}
B. Kostant, {\em The principal three--dimensional subgroup and the
Betti numbers of a complex simple Lie group.\/} \ajm{81}{1959}{973--1032}
\bibitem{LeSa}
A.N.Leznov, M.V.Saveliev, {\it Representation of zero
curvature for the system of non--linear partial differential
equations $x_{\alpha, z\bar z}=exp(kx)_\alpha$ and its integrability.\/}
\lmp{3}{1979}{489}.
\bibitem{NaRa69}
M. S. Narasimhan, S. Ramanan, {\em Moduli of vector bundles
on a compact \RS .\/} Ann. Math. {\bf 89}, (1969), 14--51.
\bibitem{RaSa93}
A. V. Razumov, M. V. Saveliev, {\em Differential geometry of Toda systems.\/}
\lanl{9311167}
\bibitem{Simp88}
C. Simpson, {\em Constructing variations of Hodge structure
using Yang--Mills theory and applications to uniformization.\/}
J. Amer. Math. Soc. {\bf 1}. (1988), 867 -- 918
\bibitem{Simp92}
C. Simpson, {\em Higgs bundles and local systems.\/} \ihes{75}{1992}{5--95}
\bibitem{SoSta91}
G. Sotkov, M. Stanishkov, {\em Affine geometry and $W_n$
algebras.\/} \np{356}{1991}{439--469}
\bibitem{Za85}
A. B. Zamolodchikov, {\em Infinite additional symmetries in two
dimensional quantum field theory.\/} Theor. Math. Phys. {\bf 63}, (1985),
1205--1210.
\bibitem{Zu92}
R. Zucchini, {\em Light cone $W_n$ geometry and its
symmetries and projective field theory.\/} \lanl{9205102}.
\end{thebibliography}
\end{document}